# Blocking Mechanism of Porn Website in India: Claim and Truth

Saurabh Pandey
Research Fellow
spandey@vmou.ac.in
Vardhman Mahaveer Open University, Kota (Rajasthan)

Dr. Harish Sharma
Associate Professor, Department of Computer Science
Rajasthan Technical Univeristy, Kota (Rajasthan)
hsharma@rtu.ac.in

## Abstract

In last few years, the addiction of internet is apparently recognized as the serious threat to the health of society. This internet addiction gives an impetus to pornographic addiction because most of the pornographic content is accessible through internet. There have been ethical concerns on blocking the contents over internet. In India Uttarakhand High court has taken initiative for the blocking of pornographic content over internet. Technocrats are coming up with various innovative mechanisms to block the content over internet with various techniques, although long ago in 2015. The Supreme Court of India has already asked to block some of the websites but it could not be materialized. The focus of this research paper is to review the effectiveness of blocking existing web content blocking mechanism of pornographic websites in Indian context.

**Introduction**

The study on the effectiveness of the blocking mechanism is crucial in knowing the current practices used for the blocking the websites in India. The report released by Kantar IMRB ICUBE 2018 reveals, that internet user base in India has crossed 500 million watermarks and is about to reach 627 million by the end of 2019. The number of internet users is suppose to be 566 million as of December 2018, registering annual growth of 18% (Pti., 2019). The following diagram presents the comparative growth of internet users since last three decades in India and China.

**Figure – 1 Internet users in India 34.4% of the population (2017)**

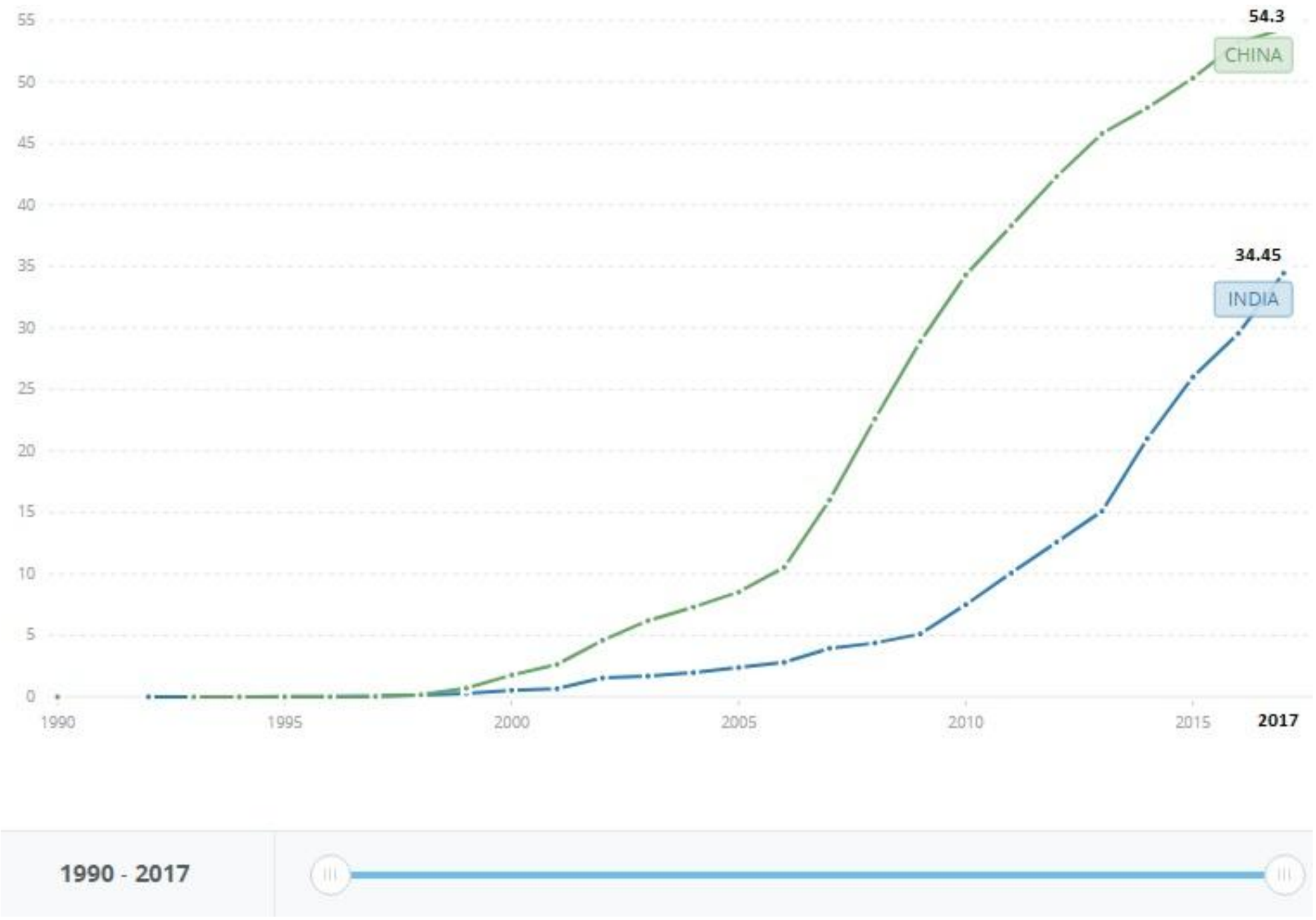

Sources:https://data.worldbank.org/indicator/IT.NET.USER.ZS?contextual=default&end=2017&locations=IN-CN&start=1990&view=chart

The rapid growth of internet users in India has made it a country prone to abuses of internet too. Kids as well as adults are gradually becoming more addictive to internet in general and unsolicited content available over internet in particular, creating a worry to the nation to block unsolicited content available over the internet to save Indian socio-cultured structure and innocence of kids. Stride efforts has been being made by Government of India (GOI) to block or restrict open access of such unsolicited content.

In one of its decision, Uttarakhand High Court states "Unlimited access to these pornographic sites is required to be blocked / curbed to avoid adverse influence on the impressionable mind of the children". The court also directed the GOI to suspend the licences of internet service providers, under Section 25 of the Information Technology Act, 2000, if they don't comply with the notification of July 31, 2015. The notification listed over more than 800 websites and directed internet service providers to block access to them as the contents posted on these websites "infringed morality and decency" (Santoshi, 2018).

The minutes of the Cyber Regulation Advisory Committee meeting held on 5th September, 2014 in DietY. Secretary, Department of Information and Technology (DOIT) informed that blocking of websites has been implemented through ISPs immediately when orders were received for blocking. The infrastructures at ISPs need to be upgraded to deal with such large number of web sites for blocking.

**Review of Related Literature**

At present, the government delegates the censorship of internet traffic through ISPs. As per the present data & facts of blocking mechanism of websites is not much encouraging then how the government might enforce a unified censorship policy for the whole county in future (Gosain *et al.*, 2018).

**Figure – 2 Growth of mobile traffic share in Pornhub's top five markets**

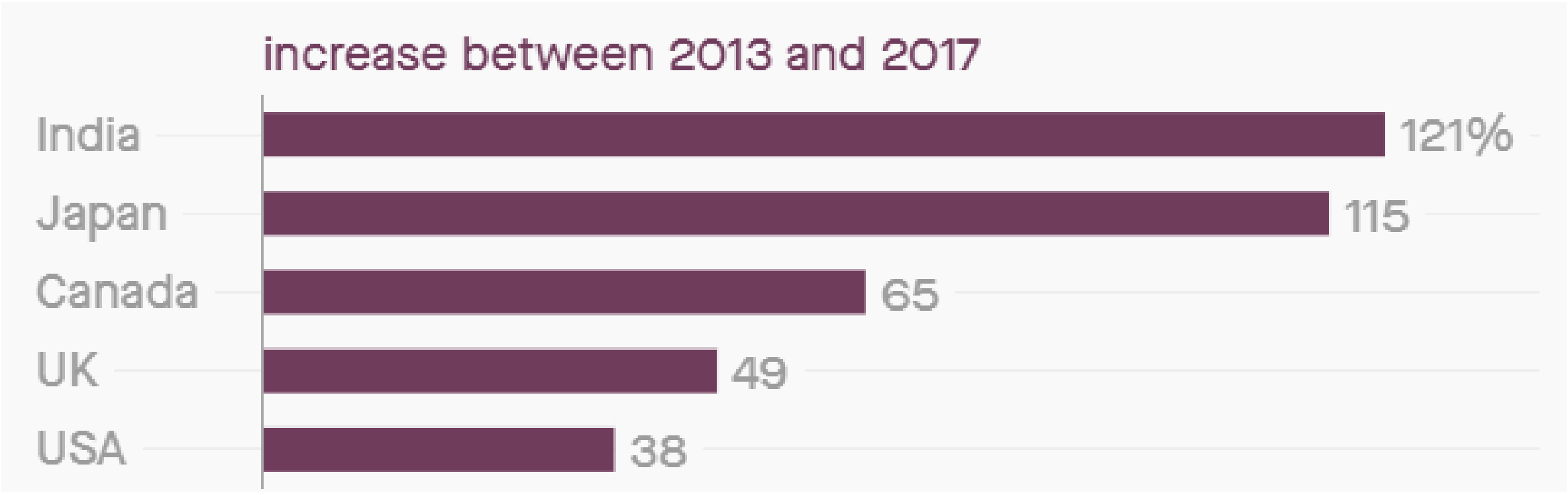

Source: Pornhub (https://www.theatlas.com/charts/HkJBO-PHG)

The viewing of adult content soared 75% since data rates came crashing down in the second half of last fiscal 2016 under severe competitive pressure, according to video viewership tracker Vidooly's findings, available exclusively with ET Telecom.com from the Economic Times. The new entrant Reliance Jio's aggressive pricing forced incumbents Bharti Airtel, Vodafone, and Idea Cellular to slash down data rates, even as porn viewing surged - primarily in the tier 2 and tier 3 towns. It is more interesting to note that about 80% of the web content is in short form, and tier 2 and 3 towns contribute 60% of total viewership

At one side research findings support the positive correlation between video consumption and its impact on brain to learn and remember things very quickly while the other side there is open internet for all age groups with the factious visuals. At present scenario, it would be the matter of big importance to differentiate between the reality of visual content against the fake and fictions ones.

Studies related to trends in pornography preferences, Pornhub reports on the device types used by its customers. Its 2018 report has shown that customers are increasingly enjoying their porn material on the go using their mobile phones. In 2013, only 40% of Pornhub's traffic was on a phone; which increased to 67% in 2017 (Kopf & Kopf 2018).

**Figure – 3 The trend of porn users moving to mobile devices**

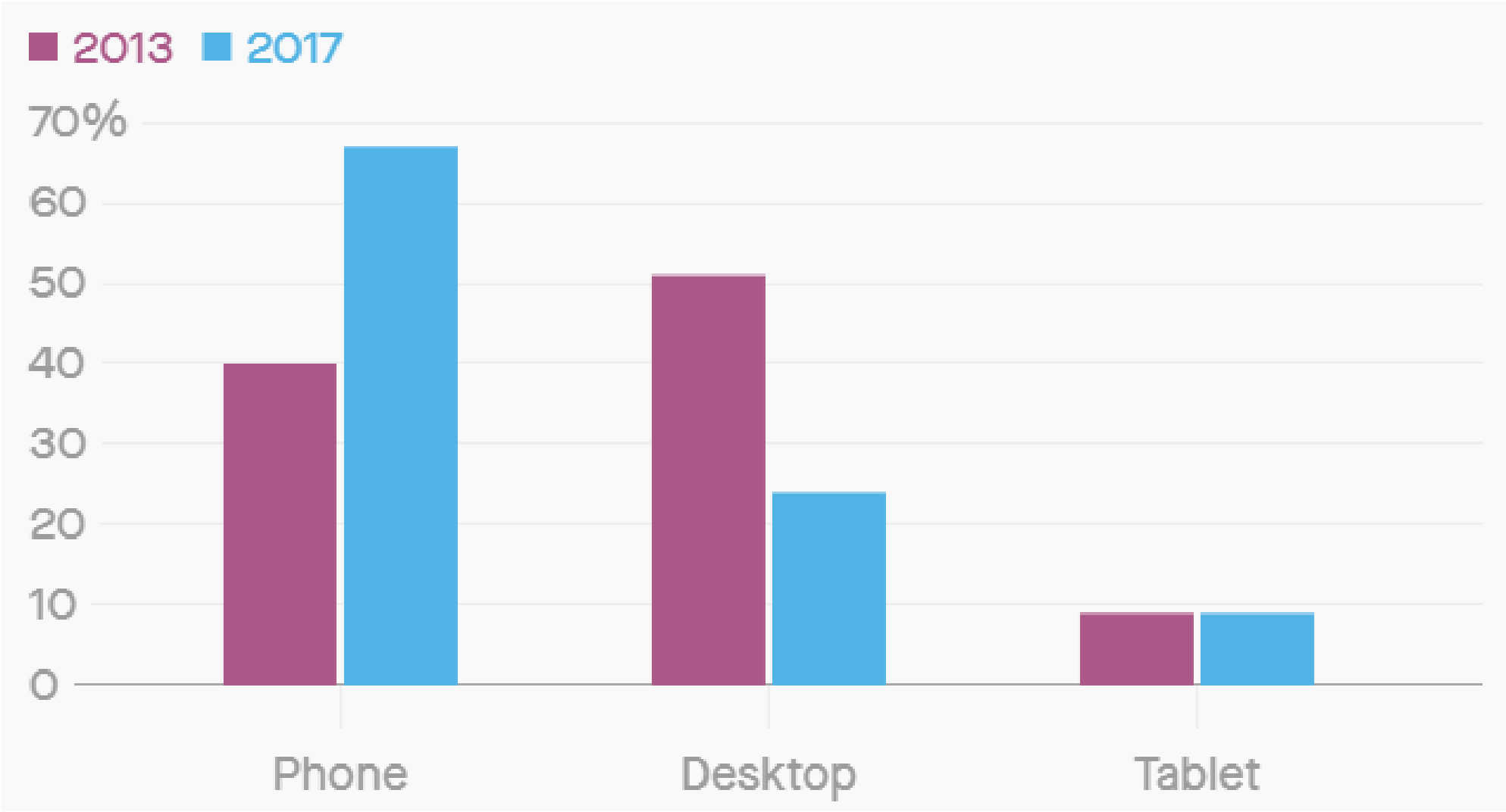

Source: Pornhub (https://www.theatlas.com/charts/BJBi_GNBM)

In recent years, internet addiction has become more prevalent worldwide and its adverse impact on the health of the society is also not invisible. Hou *et al.*, (2012) has conducted his study titled "Reduced Striatal Dopamine Transporters in People with Internet Addiction Disorder". The study revealed that internet addiction may induce significant Dopamine transporter losses in the brain and findings suggested that internet addiction is associated with dysfunctions in the dopaminergic brain systems. The findings also supported the claim that internet addiction may share similar neurobiological abnormalities with other addictive disorders.

Alarcón *et al*., (2019) have conducted a review on "Online Porn Addiction: What We Know and What We Don't". Authors concluded that pornography addiction is a type of hypersexual disorder and may be composed of several sexual behaviors, such as problematic use of online pornography (POPU). They further identified that online pornography use is on the rise, with a potential for addiction and this problematic use has adverse effects in sexual development and sexual functioning, particularly among the young population.

**Figure – 4 How the Internet of things is changing the World around us**

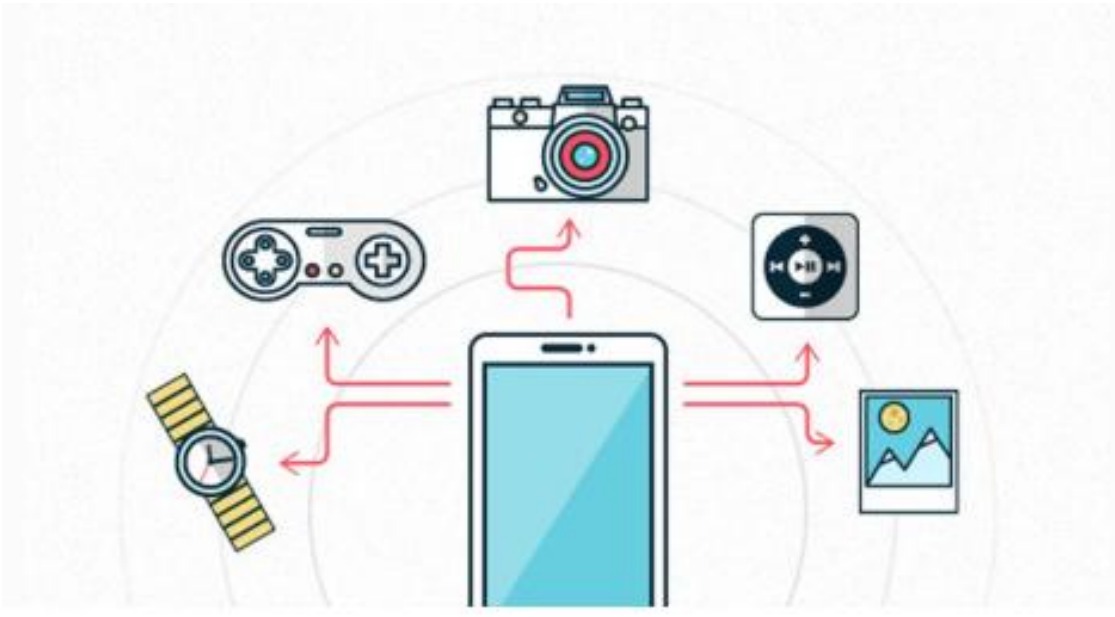

Source: https://www.netsolutions.com/insights/how-the-internet-of-things-is-changing-the-world-around-us/

In Oct. 2018, the Indian government's telecom department communicated internet service providers (ISPs) to ban 827 websites for hosting pornographic content. (Singh & Singh 2018) reported that this is the second attempt in recent times by India, among the most prolific consumers of porn watching, to shut it out. In August 2015, following a Supreme Court verdict, the government had unsuccessfully tried to block about 857 websites on the grounds that such content promotes sexual assault. Further the Uttarakhand high court has reinstated the Supreme Court's ban after a rape accused in the state's capital Dehradun said third person the culprit was nudged into committing the crime after watching a porn movie (Singh & Singh 2018).

A few years back, various websites were blocked in totality by service providers when they were only directed to block a particular webpage. The logic and rationale given by these providers was that since they lacked the intrinsic mechanism to block a webpage, they blocked the entire website.

As of now whatever are discussed were focused on abuses of internet but can be very effective if used properly for example as far as teaching and learning process is concerned watching the educational videos may be the best way to improve learning, especially when it comes to remembering key facts and figures. In fact, according to Dr. James McQuivey of Forrester Research, one minute of online video equates to approximately 1.8 million written words. In addition to this, 90% of information transmitted to the brain is visual, and visuals are processed 60,000 times faster in the brain than text. This indicates visual education aids like video can improve learning and increase the rate at which one retains the information.

There are sufficient proofs that teaching online educational videos available on *YouTube* or any other platform is facilitating the teaching and learning process. The teachers and the student both get benefited from this. The study conducted by Elyas & Kabooha, entitled "The Impacts of Using *YouTube* Videos on Learning Vocabulary in Saudi EFL classrooms" indicates that videos on you tube is making teaching and learning process attractive, interesting and enjoyable. These videos plays crucial role in motivating students intrinsically and as well as extrinsically. Alwehaibi (2015) concluded in his research that the use of *YouTube* platform resulted in reading, writing, analyzing, interacting seeking part in different activities throughout the learning cycle.

The study conducted by FSG organization a statewide pilot of Khan Academy in Idaho with 173 teachers and 10,500 students during the 2013-14 school year shows that students who complete 60% of their grade-level math on Khan Academy experience 1.8 times their expected growth on the Northwest Evaluation Association (NWEA) Measures of Academic Progress (MAP) test, which is a popular achievement of mathematics (Phillips *et al*., 2013-2014)

**Figure – 5 Impact of Khan Academy on the level of achievement of Mathematics**

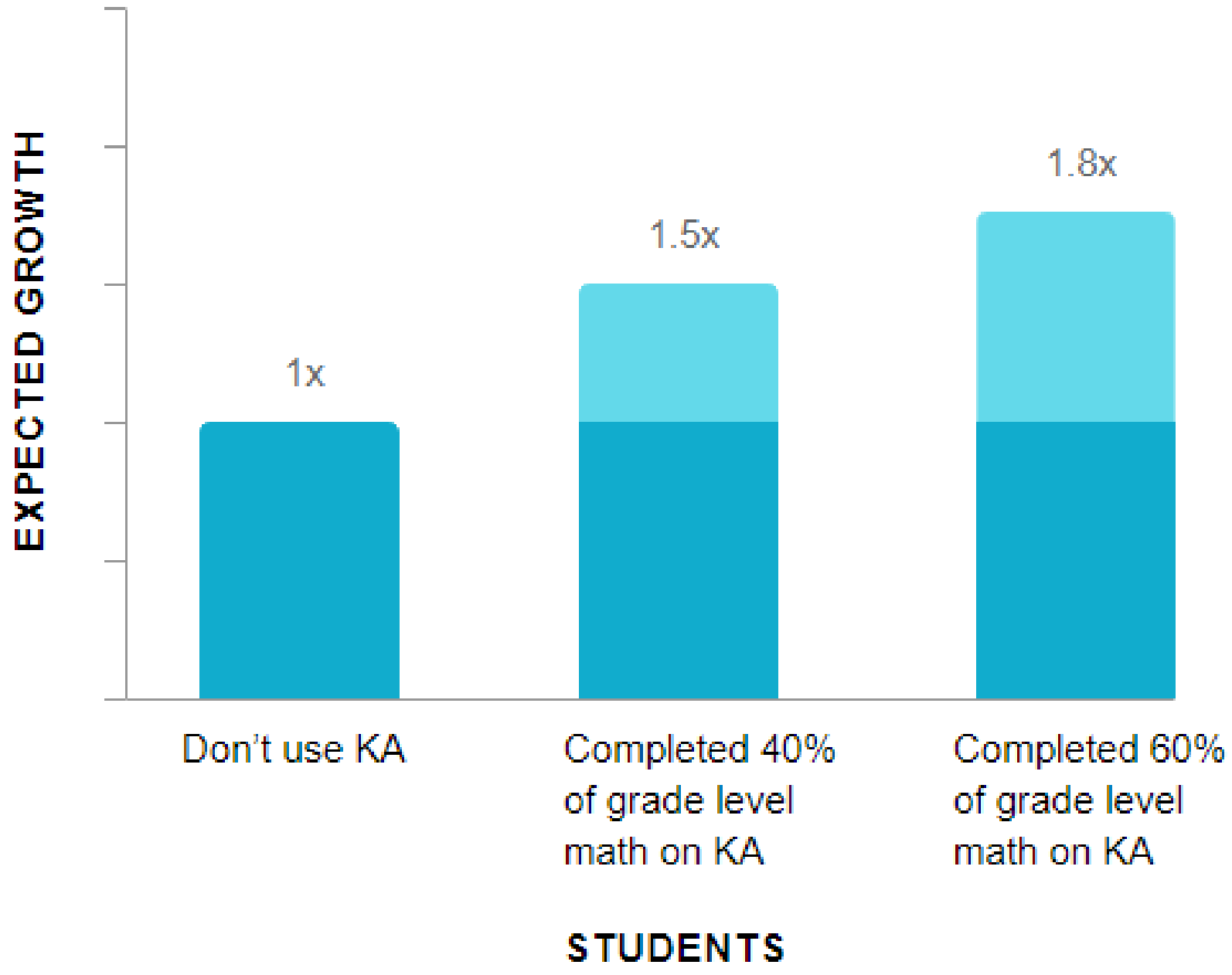

Source: https://www.khanacademy.org/about/impact

**In the present study the following objectives were under taken O1: O2:**

O1: To review the present blocking mechanism used for blocking the websites in India

Q2: To identify and analyze the Architecture of the present blocking mechanism used for blocking websites in India

In order to estimate the status of the blocked websites, the status of all the websites banned by the department of information and technology in respect of the HTTP Error Codes/ Status Codes were investigated. Using HEADMasterSEO tools 849 blocked websites were evaluated. The top four broad categories of the websites on the basis of the further higher number of counts for HTTP error code/ status code responses (200, 301, 302, blank) were identified. The HTTP status codes indicated whether a specific HTTP request has been successfully completed or not. Every HTTP request that is received by a server is responded with an HTTP status code. HTTP status codes consist of three digit codes and grouped in different classes. The class of status code identified by its first digit of the status code responses –

1xx: Informational

2xx: Success

| **Table – 1 Description of Error Code/ Status Code 200** | | |
|---|---|---|
| Error Code / Status Code | Message | Description |
| 200 | OK | This code indicates that server successfully processed the request and in response provided the requested page. |

3xx: Redirection

| **Table – 2 Description of Error Code/ Status Code 301, 302** | | |
|---|---|---|
| Error Code / Status Code | Message | Description |
| 301 | Moved Permanently | This code refers that the requested page permanently moved to a different location. It means server forwards the user request to the new associated location. |
| 302 | Found / Moved Temporarily | This code refers that presently server is catering the user request from the different location, but in future the user request may be catered from the previous or original location. |

4xx: Client Error

5xx: Server Error

| **Table - 3 An overview of Technical Status of blocked websites in India based on Error Codes** | | | | | | | | |
|---|---|---|---|---|---|---|---|---|
| Error Code Description | Error Codes | | | | | | | No of URL |
| Row Labels | 200 | 301 | 302 | 307 | 403 | 404 | (blank) | Grand Total |
| Forbidden | | | | | 3 | | | 3 |
| Found | | | 43 | | | | | 43 |
| Found : Moved Temporarily | | | 1 | | | | | 1 |
| Found, Redirect Loop | | | 3 | | | | | 3 |
| Invalid Server Response | | | | | | | 10 | 10 |
| Moved Permanently | | 186 | | | | | | 186 |
| Moved Temporarily | | | 6 | | | | | 6 |
| Name Not Resolved | | | | | | | 137 | 137 |
| Not Found | | | | | | 4 | | 4 |
| OK | 449 | | | | | | | 449 |
| Redirect | | | 2 | | | | | 2 |
| Temporary Redirect | | | | 4 | | | | 4 |
| Timeout | | | | | | | 1 | 1 |
| Grand Total | 449 | 186 | 55 | 4 | 3 | 4 | 148 | 849 |

| Table - 4 Top Four Broad Categories of Websites based on Error Codes | | | | | |
|---|---|---|---|---|---|
| | Error Codes | | | | No of URL's |
| Description : Error Code | 200 | 301 | 302 | 0 (blank) | Grand Total |
| Found | | | 43 | | 43 |
| Moved Permanently | | 186 | | | 186 |
| Name Not Resolved | | | | 137 | 137 |
| OK | 449 | | | | 449 |
| Grand Total | 449 | 186 | 43 | 137 | 815 |

**Sample and Sampling**

In this study 262 porn websites were selected as sample through proportionate stratified random sampling method. The sample size was determined using formula for determination of sample size for known population (Cochran, 1963, p. 75).

Sample size for population $$\frac{n_0 = z^2 * P(1-P)}{e^2}$$ Equation 1

Where

$n_0$ = Initial sample size

z = Selected critical values of desired level of confidence or risk from (Z-table) 90% (1.645), 95% (1.96), 99% (2.576)

P = Estimated proportion of an attribute that is present in the population of maximum variability of the population Maximum Variability (50%) that is (0.5)

e = Desired level of precision or margin of error (+- 5%) that is (0.05)

Now putting the values in formula

$$\frac{n_0 = 1.96^2 * 0.5(1-0.5)}{(0.05)^2}$$

$$= 384.16 = 384$$

Now calculate population correction factor for sample size for known population –

$$n = \frac{n_0 N}{n_0 + (N-1)}$$ Equation 2

Where

n = Sample size for known population

$n_0$ = Initial sample size for population

N = Known population

$$n = \frac{384 * 815}{384 + 814}$$

$$= \frac{312960}{1198}$$

$$= 261.23 = 261$$

Sample size of the strata = $\frac{Size\ of\ entire\ sample}{population\ size\ *\ layer\ size}$ Equation 3

Now putting the values in formula

| Table – 5 Sample Description | | | |
|---|---|---|---|
| Strata | Known Population | Calculations | Sample |
| 200 | 449 | $\frac{261}{815 * 449} = 143.79 = 144$ | 144 |
| 301 | 186 | $\frac{261}{815 * 186} = 59.56 = 60$ | 60 |
| 0 | 137 | $\frac{261}{815 * 137} = 43.87 = 44$ | 44 |
| 302 | 43 | $\frac{261}{815 * 43} = 13.77 = 14$ | 14 |
| | | | 262 |

**Experimental Results: Analysis and Discussion**

When attempt were made to access the blocked website / URL's to use of the tools (tor browser and opera browser with virtual private network (VPN) facility) which uses the different ports and protocol of the Open System Interconnection (OSI) model for the communication to serve the user request. The end results were very awaking that more than 95% of the website in each strata were successfully browsed with these tools and technology except the HTTP 0/Blank error codes/ status codes (Name Not Resolved) websites. Name Not Resolved means that the hostname/ website you are trying to connect cannot be resolved to an IP address (Hostname/ website name resolved to IP addresses by a Domain Name Server system).

| Table - 6 Tor Browser Success % | | |
|---|---|---|
| Error Code | Description | Tor Browser Success % |
| 302 | Found | 96 |
| 301 | Moved Permanently | 98 |
| 200 | OK | 99 |

**Figure - 6 Tor Browser Success %**

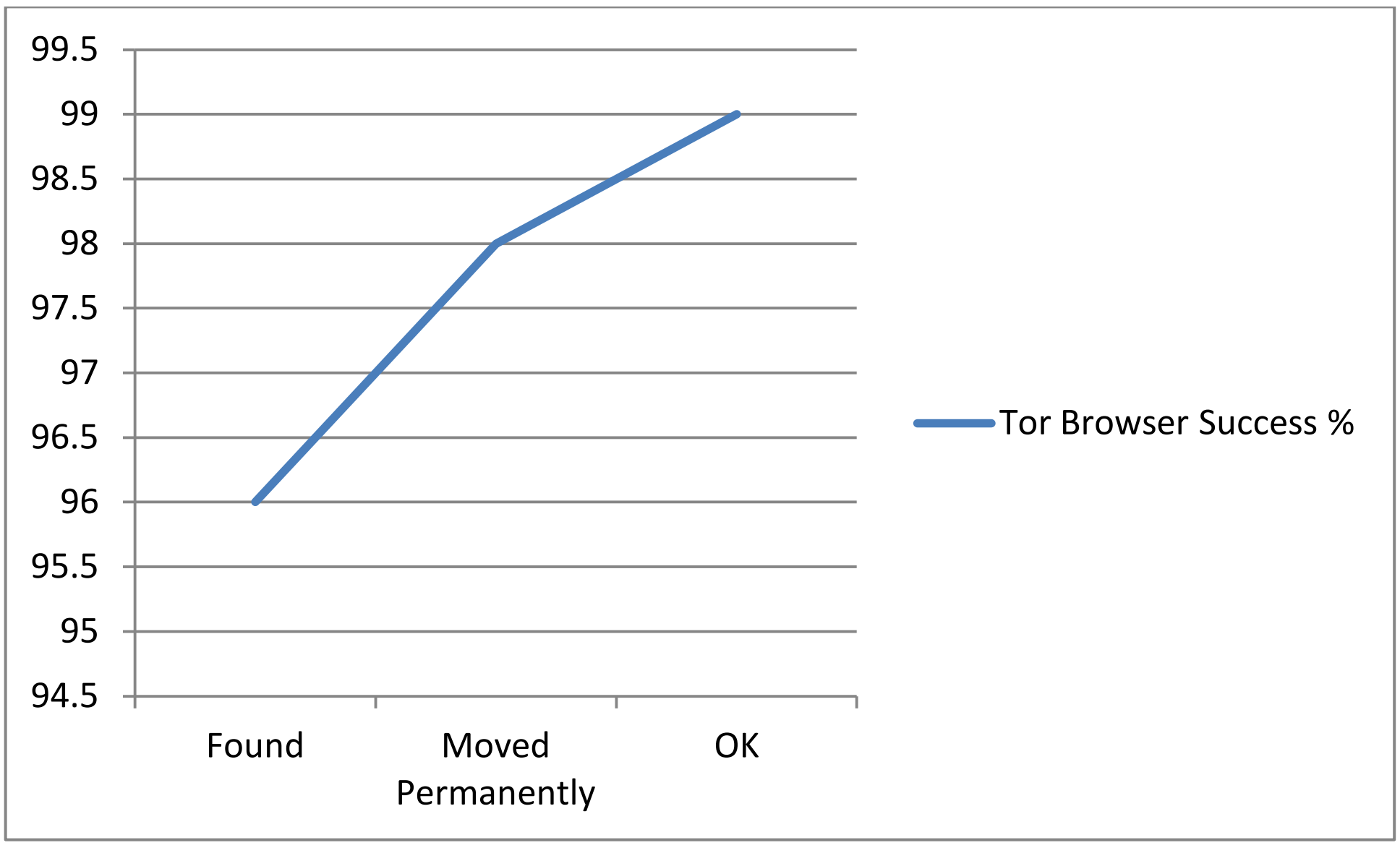

Table 6 and Figure 6 represents that porn websites across different category of error codes/ response codes (302, 301 and 200), are brows able with Tor browser at the success rate of more than 96%.

| **Table - 7 Opera Browser (VPN) Success %** | | |
|---|---|---|
| Error Code | Description | Opera Browser (VPN) Success % |
| 302 | Found | 97 |
| 301 | Moved Permanently | 96 |
| 200 | OK | 98 |

**Figure 7 Opera Browser (VPN) Success in percentage**

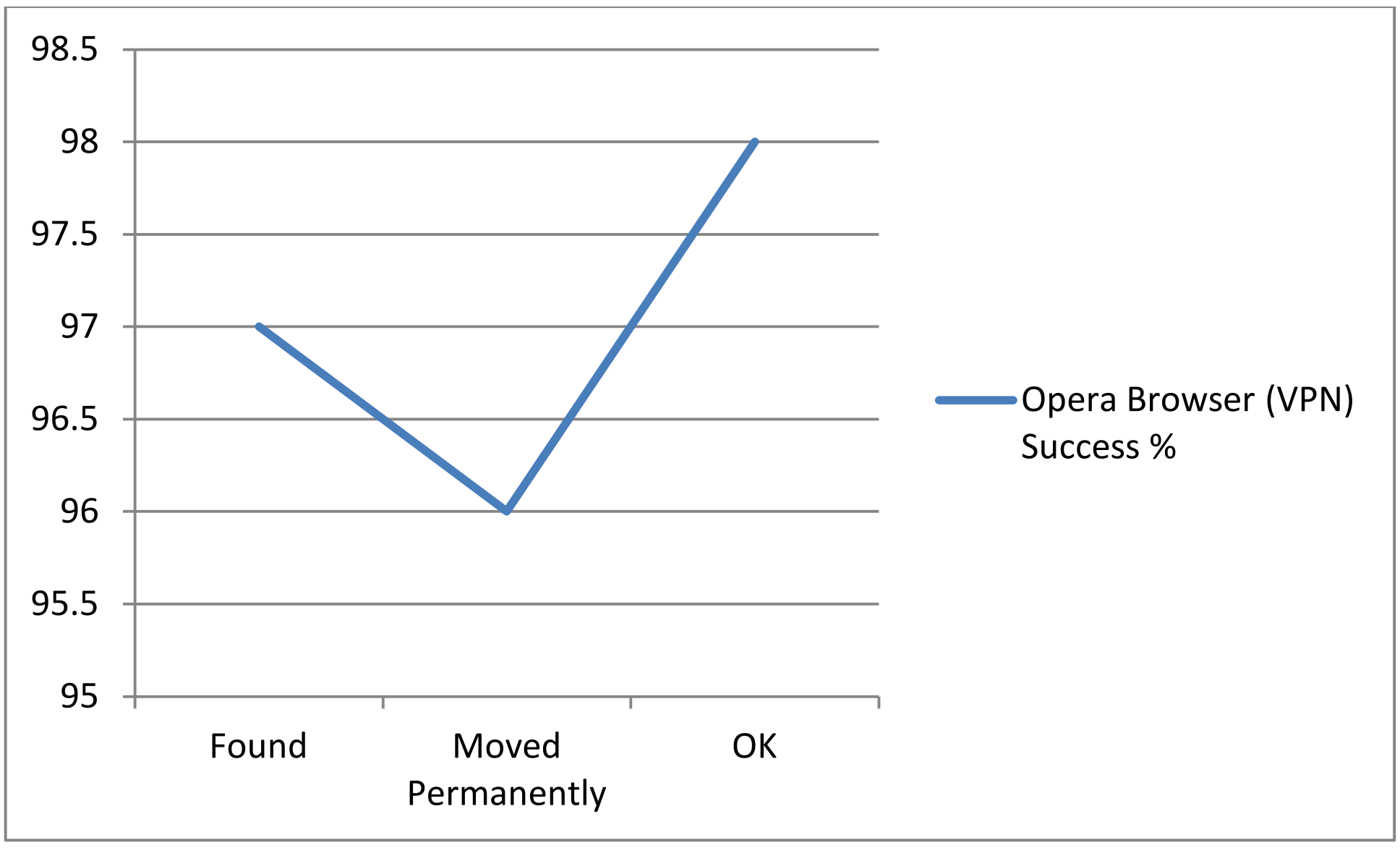

Table 7 and Figure 7 represent that porn websites under the category of error code/ response code (302, 301 and 200) are brows able with Opera browser at the success rate of more than 96%.

| **Table - 8 Comparison of Tor Browser & Opera Browser (VPN) Success %** | | | |
|---|---|---|---|
| Error Code | Description | Tor Browser Success % | Opera Browser (VPN) Success % |
| 302 | Found | 96 | 97 |
| 301 | Moved Permanently | 98 | 96 |
| 200 | OK | 99 | 98 |

**Figure - 8 Comparison of Tor Browser & Opera Browser (VPN) Success in percentage**

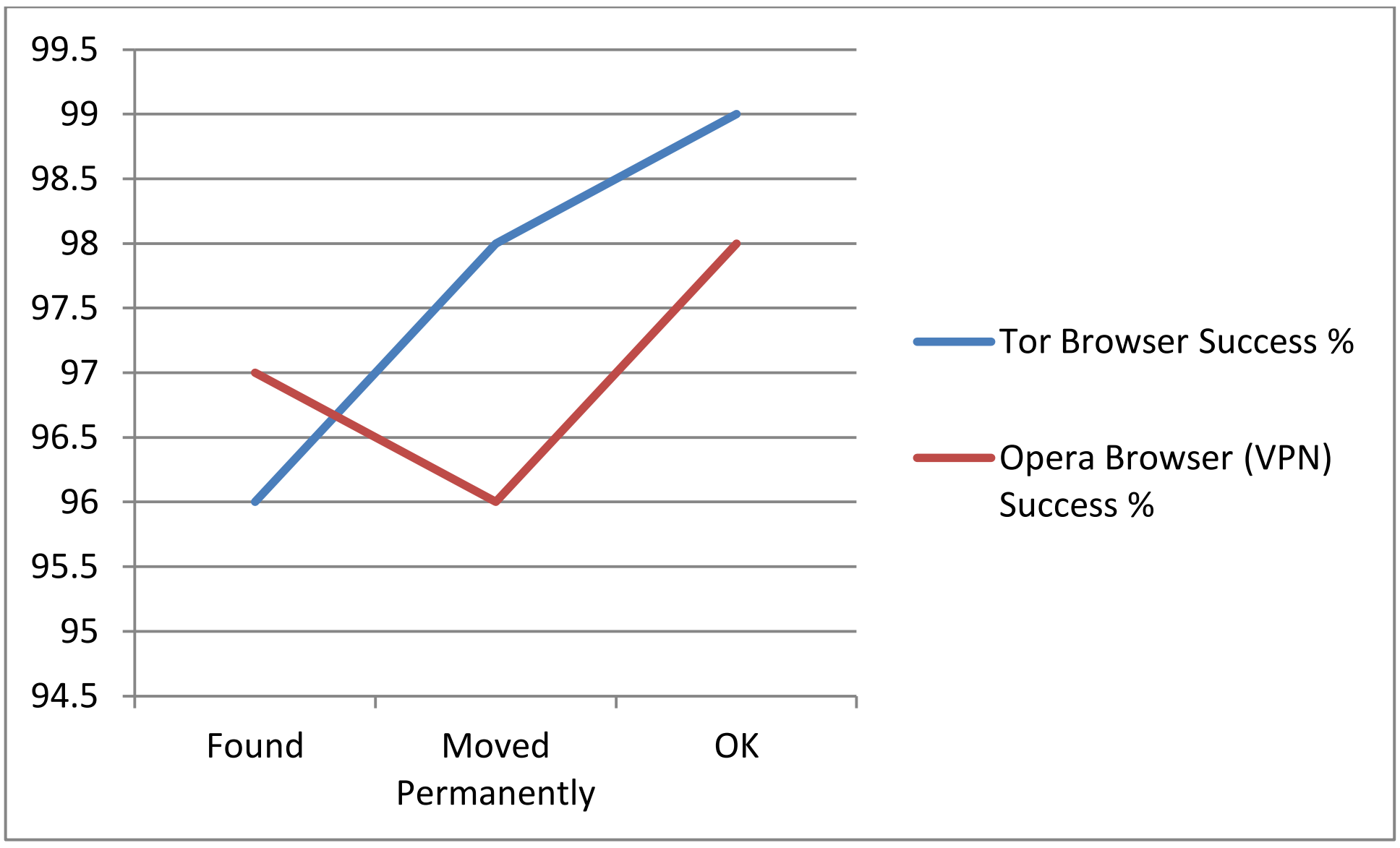

Table 8 and Figure 8 indicate that more than 95% of the websites were selected as sample is easily accessible with the help of these browsers.

| Table - 9 Status of Websites with Blank Error Codes | |
|---|---|
| Status of websites with blank error codes | No. of Domains |
| Domain Name Available for Purchase | 24 |
| A Records Not Found | 21 |
| Total No of Websites | 45 |

**Figure – 9 Status of Website with Blank Error Codes**

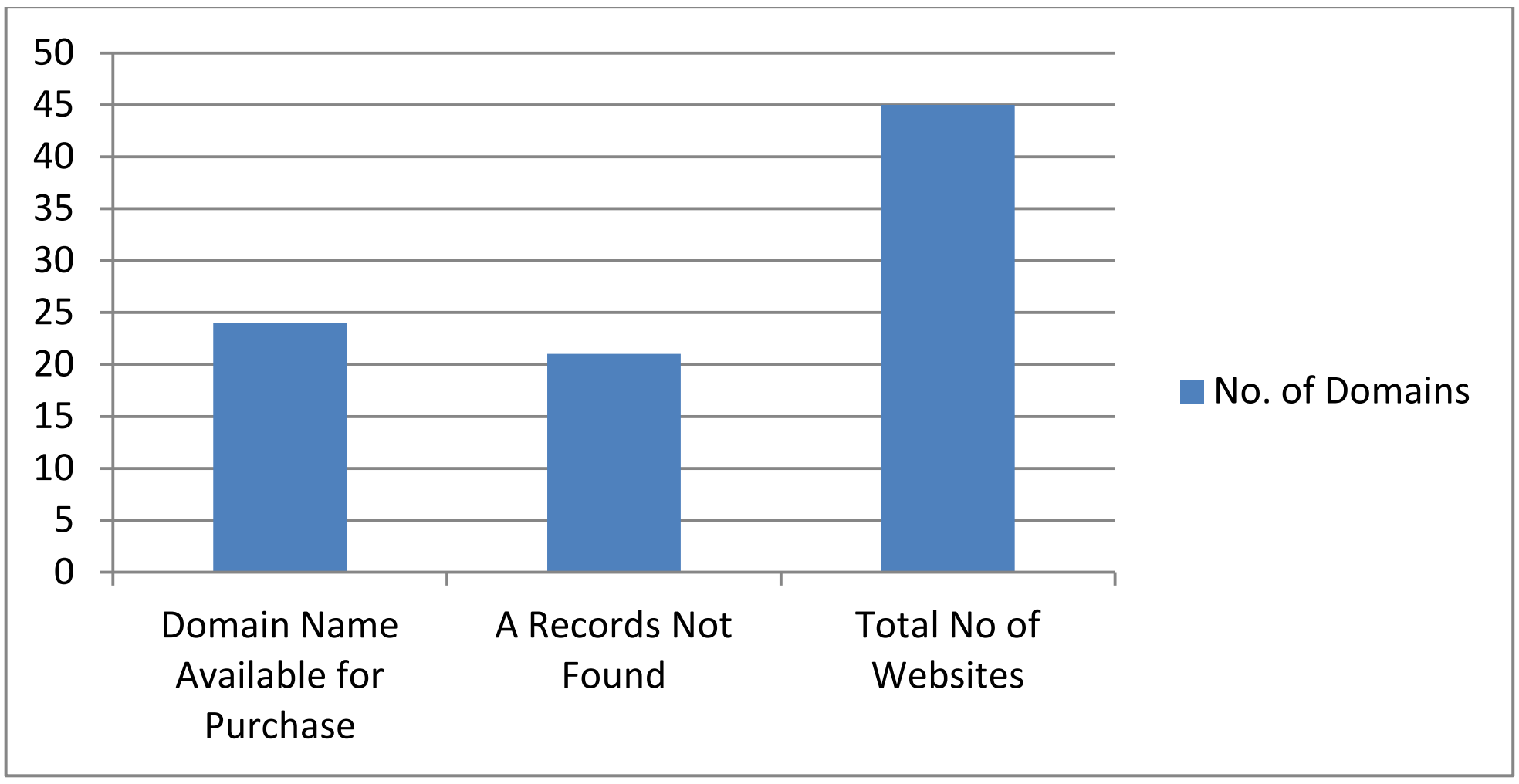

Table 9 and Figure 9 show the status of the domain name of the websites for the bank error codes. Out of 45 websites in this category 24 domain names available for purchase for this testing, were used the godaddy bulk domain check web utility Register Domains in bulk at GoDaddy. 21 domain names of websites existed but not mapped with any A records for this testing we have used the infobyip.com web utility page ipbulklookup.php (InfoByIP.com.).

**Findings -** The testing results reveals that the DNS based content blocking mechanisms are being used by the ISP's in India. Website's blocking has been done in continuation to the blocking orders & list released from the department of information and technology (DOIT) in compliance of the order to ban websites by Uttarakhand High Court. This type of filtering mechanism caters the user DNS requests, either by dropping the request or by responding the misleading IP responses by the DNS. Although, in India there are approx 50000 DNS servers, distributed across different networks, reconfiguring all such servers to filter DNS queries for black-listed sites would not be easy (Gosain *et al.*, 2018). DNS based content blocking does not filter and examine all network traffic; instead, it focuses on controlling and routing of DNS queries to bogus IP's.

**Discussion on Findings –** Checking of the status of the A records for the blocked websites in India and from the US locations through the help of the online utility available at InfoByIP.com, revealed the same A records values from both the locations. Further when it was tried to access the banned website in India, it routes the request and finally got the block page of the website with the text message (Web Page Blocked! The page you have requested has been blocked, because the URL is banned as per the Government Rules) and the meanwhile when accessed the same website with United States hosted server it lead to the final destination of the website without any block page message because there is no any restriction in United States for these websites. As per these testing and comparison of the results for two different geographical locations lookup results for A records. End result indicated that for the same public IP (A public IP address is the globally unique IP address assigned to any device) one is getting two different page responses due to the DNS based content blocking in India. In the United States the web request with the same IP lead to the actual destination of the web page while in India it routes to the blocked page hosted / routed by the various ISP's in India.

So on the bases of above experimental results one can easily interpret that the ISP's of India often uses the DNS based content blocking mechanism to block the websites in India.

**Figure - 10 Architecture of DNS-Based Websites Blocking Mechanism**

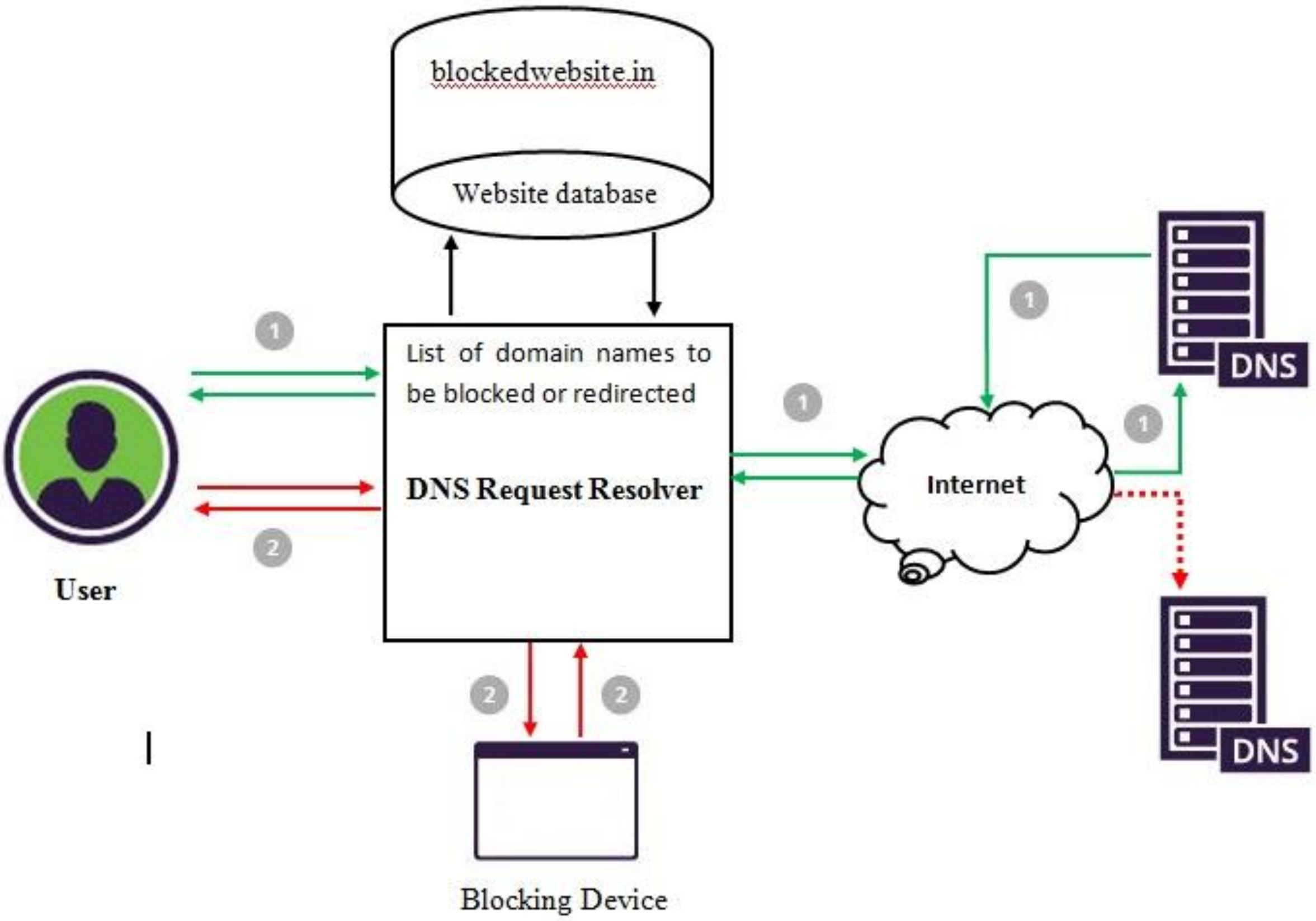

User initiates the request for website i.e. www.allowwebsite.in The allowwebsite.in is not in the list of blocked domain names. User request forwarded to the DNS server for allowwebsite.in with the query what is the IP address of allowwebsite.in

DNS server for allowwebsite.in resolves the request and in response send the IP address of allowwebsite.in=W.X.Y.Z to the user. The DNS query operate normally and returned the correct answer

User initiates the request for website i.e. blockedwebsite.in The blockedwebsite.in lies in the database of blocked website list. In this case the user request will not forwarded to the DNS server

The DNS query for blockwebsite.in intercepted by the blocking device which returns the IP address of the webserver. Webserver returns the error page “Website Blocked” because the blockwebsite.in on the block list

**Conclusion**

The present blocking mechanism of the Government of India seems not to be effective are to address the issue with the long lasting corrective measures. The present blocking mechanism is not sustainable to achieve the practical goals at grass root level. The present blocking mechanism used by the ISP's is low cost and can be quickly implemented, having the major flaw not to deal with the provision of blocking mechanism at the different layers and ports of the OSI / TCPIP model. The present generation of the current decade is very friendly to computer and use of internet. The present blocking mechanism is not holistic solution to block the request for the unwanted sites at different levels. In present study it is observed that the blocked websites are easily accessible with the help of the browsers, using the different protocols and ports of communication. As shown in table no – 9 the data of the domain names available for purchase is 24 means that only 24 blocked websites shutdown with the blocked domains and may be migrated with the other domain names easily. There are numerous pornographic websites available on internet so it does not seem practical to block it with the help of ISP's only using DNS based blocking. India has to think of an integrated holistic model for web content filtering and blocking.

**References**

Aggarwal, A. (2017, July 10). *How the Internet of Things is changing the World around Us*. Retrieved from https://www.netsolutions.com/insights/how-the-internet-of-things-is-changing-the-world-around-us/

Alarcón, R. D., Iglesia, J. D., Casado, N., & Montejo, A. (2019). Online Porn Addiction: What We Know and What We Don't—A Systematic Review. *Journal of Clinical Medicine,8*(1), 91. doi:10.3390/jcm8010091

Alwehaibi, H. O. (2015). The Impact Of Using *YouTube* In EFL Classroom On Enhancing EFL Students Content Learning. *Journal of College Teaching & Learning (TLC), 12*(2), 121. doi:10.19030/tlc.v12i2.9182

Gosain, D., Agarwal, A., Shekhawat, S., Acharya, H. B., & Chakravarty, S. (2018). *Mending Wall: On the Implementation of Censorship in India*. [Lecture Notes of the Institute for Computer Sciences, Social Informatics and Telecommunications Engineering Security and Privacy in Communication Networks], 418-437. doi:10.1007/978-3-319-78813-5_21

Here is the full list of 827 porn websites blocked by DoT. (2018, December 15). Retrieved January 24, 2019, from https://indianexpress.com/article/technology/tech-news-technology/here-is-the-full-list-of-827-porn-websites-banned-by-the-dot-5421127/

Hou, H., Jia, S., Hu, S., Fan, R., Sun, W., Sun, T., & Zhang, H. (2012). Reduced Striatal Dopamine Transporters in People with Internet Addiction Disorder. *Journal of Biomedicine and Biotechnology,2012*, 1-5.

InfoByIP.com. (n.d.). *Domain and IP bulk lookup tool*. Retrieved May 19, 2019, from https://www.infobyip.com/ipbulklookup.php

Kopf, D., & Kopf, D. (2018, December 13). *Forget Netflix-Pornhub tells us everything we need to know about the future of internet viewing habits*. Retrieved January 11, 2019, from https://qz.com/1186286/data-show-porn-is-moving-to-mobile/

Livemint. (2019, March 11). *India's internet base crosses 500 million mark, driven by Rural India*. Retrieved April 3, 2019, from https://www.livemint.com/industry/telecom/internet-users-exceed-500-million-rural-india-driving-growth-report-1552300847307.html

Phillips, D., & Cohen, J. (n.d.). *Impact*. Retrieved August 17, 2018, from https://www.khanacademy.org/about/impact

Pti. (2019, March 06). *Internet users in India to reach 627 million in 2019: Report*. Retrieved April 21, 2019, from https://economictimes.indiatimes.com/tech/internet/internet-users-in-india-to-reach-627-million-in-2019-report/articleshow/68288868.cms

Register Domains in bulk at GoDaddy. (n.d.). Retrieved May 19, 2019, from https://in.godaddy.com/domains/bulk-domain-search.aspx

Santoshi, N. (2018, September 28). *Unlimited access to porn sites should be curbed: Uttarakhand high court*. Retrieved December 9, 2019, from https://www.hindustantimes.com/dehradun/unlimited-access-to-porn-sites-should-be-curbed-uttarakhand-high-court/story-3xBQ8yWjU9rTknhXw43wYN.html

Singh, K., & Singh, K. (2018, November 30). *India is trying to ban porn again. Here's why it will fail*. Retrieved December 15, 2018, from https://qz.com/india/1441110/how-indians-still-visit-pornhub-despite-the-porn-ban/

Tsur, M. (2014, June 01*). Research Confirms Video Improves Learning Results*. Retrieved August 24, 2017, from https://www.huffingtonpost.com/michal-tsur/research-confirms-video-i_b_5064181.html

Www.ETTelecom.com. (2017, June 02). *Porn viewing on smartphones up 75% as data rates drop in India - ET Telecom*. Retrieved January 21, 2018, from https://telecom.economictimes.indiatimes.com/news/porn-viewing-on-smartphones-surges-75-as-data-rates-drop-in-india/58966755

**Annexure A**

## List of Total (815) URL's/ Websites taken as known population

1 http://theporndude.com/
2 http://indianporntube.xxx/
3 http://sexsex.hu/
4 http://hot-dates.info/
5 http://en.cam4.com.br/
6 http://videos.petardas.com/
7 http://es.porn.com/
8 http://jav-porn.net/
9 http://callboyindia.com/
10 http://pornfromczech.com/
11 http://mypornbookmarks.com/
12 http://bomnporn.com/
13 http://africansexvideos.net/
14 http://kirtu.com/
15 http://eroticperfection.com/
16 http://hqlinks.net/
17 http://dslady.com/
18 http://gaypornium.com/
19 http://en.cam4.co/
20 http://indiansex4u.com/
21 http://teengayporntube.com/
22 http://sexxxxi.com/
23 http://3rat.com/
24 http://4hen.com/
25 http://luboeporno.com/
26 http://shemale.asia/
27 http://thefreecamsecret.com/
28 http://fapto.xxx/
29 http://gracefulnudes.com/
30 http://cam4.in/
31 http://cullosgratis.com.ve/
32 http://bananabunny.com/
33 http://legalporno.com/
34 http://desimurga.com/
35 http://pinkythekinky.com/
36 http://gracefulnudes.com/
37 http://ixxx.com/
38 http://indlansex.net/
39 http://momsteachsex.com/
40 http://analsexstars.com/
409 http://hotnudematures.com/
410 http://bigfreemature.com/
411 http://milfionaire.com/
412 http://wifezilla.com/
413 http://freematurevideo.net/
414 http://maturenakedsluts.com/
415 http://tgpmaturewoman.com/
416 http://mature-women-tube.org/
417 http://matureal.com/
418 http://bestmilftube.com/
419 http://dianapost.com/
420 http://hotfreemilfs.com/
421 http://momsnightjob.com/
422 http://thexmilf.com/
423 http://numaturewomen.com/
424 http://chubbygalls.com/
425 http://silkymoms.com/
426 http://riomoms.com/
427 http://booloo.com/
428 http://fuckmaturewhore.com/
429 http://hqmaturemovs.com/
430 http://sharedxpics.com/
431 http://bestmatureclips.com/
432 http://viewmature.com/
433 http://milfgals.net/
434 http://matureintros.com/
435 http://maturecool.com/
436 http://matureasspics.com/
437 http://milfmomspics.com/
438 http://maturepornpics.com/
439 http://milfpicshere.com/
440 http://mature30-45.com/
441 http://el-ladies.com/
442 http://womenmaturepics.com/
443 http://bigbuttmature.com/
444 http://mygranny.pics/
445 http://momspics.net/
446 http://cleomture.com/
447 http://secinsurance.com/
448 http://idealwifes.com/

41 http://pinkworld.com/
42 http://89.com/
43 http://indianpornovid.com/
44 http://playboy.com/
45 http://locasporfollar.com/
46 http://perfectgirls.net/
47 http://hairy.com/
48 http://fuckcuck.com/
49 http://ixxx.com.es/
50 http://darering.com/
51 http://drtuber.com/
52 http://epicporntube.com/
53 http://hindisex.com/
54 http://milfmovs.com/
55 http://sexocean.com/
56 http://pornorc.net/
57 http://thefreecamsecret.com/
58 http://teenpornxxx.net/
59 http://livjasmin.com/
60 http://naughty.com/
61 http://perucaseras.com/
62 http://xxx.com/
63 http://x-ho.com/
64 http://cliphunter.com/
65 http://watchmygf.com/
66 http://tnaflix.com/
67 http://truthordarepics.com/
68 http://xxx.com/
69 http://adultsextube.com/
70 http://xhamster.com/
71 http://freshporn.info/
72 http://xvideos.com/
73 http://xhot.sextgem.com/
74 http://ixxx.ws/
75 http://brazzers.com/
76 http://porntube.com/
77 http://hot-gifz.com/
78 http://parejasfollando.es/
79 http://purebbwtube.com/
80 http://lechecallente.com/
81 http://es.chaturbate.com/
82 http://pornhub.com/
83 http://fotomujeres.pibones.com/
84 http://jizzhut.com/
85 http://savitabhabhi.mobi/
449 http://momsclan.com/
450 http://mature.nl/
451 http://alloldpics.com/
452 http://xebonygirls.com/
453 http://milf-fucking.net/
454 http://bravomamas.com/
455 http://matureholes.net/
456 http://mamitatube.com/
457 http://matureclithunter.com/
458 http://tgpmaturewoman.com/
459 http://idealmilf.com/
460 http://mulligansmilfs.com/
461 http://hardsexyyoupornhub.com/
462 http://pretty-matures.com/
463 http://immoralmatures.com/
464 http://hotelmatures.com/
465 http://juicy-matures.com/
466 http://mturemomsporn.com/
467 http://bestmaturesthumb.com/
468 http://vipoldies.net/
469 http://maturesexy.us/
470 http://maturegirl.us/
471 http://pussy-mature.com/
472 http://galsarchive.com/
473 http://bigtitsnaked.com/
474 http://ahmilf.com/
475 http://nudemomphotos.com/
476 http://petiteporn.pw/
477 http://maturecherry.net/
478 http://sexpics.xxx/
479 http://lewd-babes.com/
480 http://nudematurepussy.com/
481 http://sexyhotmilf.com/
482 http://olderwomentaboo.com/
483 http://wethairywats.com/
484 http://alexmatures.com/
485 http://maturedally.net/
486 http://nudematuremix.com/
487 http://xxxmomclips.com/
488 http://alansanal.com/
489 http://ladymom.com/
490 http://chocomilf.com/
491 http://fresh-galleries.com/
492 http://ladymom.com/
493 http://womeninyears.com/

86 http://alohatube.com/
87 http://indianpornvideos.com/
88 http://pornochaud.com/
89 http://gokabyle.com/
90 http://rubber-kingdom.com/
91 http://fuq.com/
92 http://pornxxxtubes.com/
93 http://2gayboys.com/
94 http://porno.com/
95 http://freex.mobi/
96 http://pinkvisualtgp.com/
97 http://porn.mangassex.com/
98 http://goulnes.pornoxxxi.net/
99 http://freeones.com/
100 http://videos-x.xpornogays.com/
101 http://voyeurpipi.com/
102 http://arabebaise.com/
103 http://rk.com/
104 http://conejox.com/
105 http://xvideosnacional.com/
106 http://xxxonxxx.com/
107 http://bdenjoymore.blogspot.com/
108 http://jeux-flash-sexy.com/
109 http://iknowthatgirl.com/
110 http://ohasiatique.com/
111 http://petardas.com/
112 http://xnxx.com/
113 http://cumlouder.com/
114 http://marocainenue.com/
115 http://h33t.to/
116 http://fr.perfectgirls.net/
117 http://gayboystube.com/
118 http://lisaannlovers11.tumblr.com/
119 http://youjizz.com/
120 http://flirthookup.com/
121 http://tubegalore.com/
122 http://youporn.com/
123 http://playvid.com/
124 http://roundandbrown.com/
125 http://conejox.com/
126 http://fille-nue-video.com/
127 http://bootlus.com/
128 http://gorgeousladies.com/
129 http://videosdemadurasx.com/
130 http://videosfilleschaudes.com/
494 http://picsboob.com/
495 http://cheatwife.com/
496 http://40somethingmag.com/
497 http://hotchicks.sexy/
498 http://maturehere.com/
499 http://bigtitsmilf.com/
500 http://fatsexygirls.net/
501 http://milfsection.met/
502 http://bestmilfsporn.com/
503 http://mature-beach.com/
504 http://horny-olders.com/
505 http://momsforporn.com/
506 http://empflix.com/
507 http://sexy-olders.com/
508 http://older-beauty.com/
509 http://place21.com/
510 http://hairymaturegirls.com/
511 http://maturetube.com/
512 http://grandmammapics.com/
513 http://myexmilf.com/
514 http://gracefulmilf.com/
515 http://xmilfpics.com/
516 http://maturemompics.com/
517 http://mature-library.com/
518 http://numoms.com/
519 http://sexualolders.com/
520 http://ebonyfantasies.com/
521 http://mature4.net/
522 http://azgals.com/
523 http://milfera.com/
524 http://icematures.com/
525 http://agedmamas.com/
526 http://sexymaturethumbs.com/
527 http://erotic-olders.com/
528 http://classic-moms.com/
529 http://filthymamas.com/
530 http://excitingmatures.com/
531 http://nudematuremix.com/
532 http://milfjam.com/
533 http://sexymaturethumbs.com/
534 http://freematurepornpics.com/
535 http://bbwpornpics.com/
536 http://milfbank.com/
537 http://maturebrotherthumbs.com/
538 http://gramateurs.com/

131 http://bangbros.com/
132 http://serviporno.com/
133 http://sexxxdoll.com/
134 http://cholotube.com/
135 http://xtube.com/
136 http://xxxvideosex.org/
137 http://videosxxxputas.xxx/
138 http://teensnow.com/
139 http://babes.com/
140 http://saoulbafjojo.com/
141 http://sexonapria.org/
142 http://coffetube.com/
143 http://yourather.com/
144 http://myfreecams.com/
145 http://femmesmuresx.net/
146 http://gaygautemela.com/
147 http://couleurivoire.com/
148 http://lesbiennesxxx.com/
149 http://beurettehot.net/
150 http://redtube.com/
151 http://videos-porno-chaudes.com/
152 http://3animalsextube.com/
153 http://fillechaude.com/
154 http://xgouines.com/
155 http://premiercastingporno.com/
156 http://pornofemmeblack.com/
157 http://poringa.net/
158 http://freesex.com/
159 http://woodstockreborn.tumblr.com/
160 http://porno-algerienne.com/
161 http://moncotube.net/
162 http://sexcoachapp.com/
163 http://awesomeellalove.tumblr.com/
164 http://ixxx-tube.com/
165 http://sexocean.com/
166 http://des-filles-sexy.com/
167 http://top-chatroulette.com/
168 http://babosas.com/
169 http://femdomecpire.com/
170 http://tube8.com/
171 http://pornmotion.com/
172 http://videos-sexe.1touffe.com/
173 http://tubeduporno.com/
174 http://xnxx-free.net/
175 http://xxi.onxxille.com/
539 http://thematureladies.com/
540 http://eroticteens.pw/
541 http://pamelapost.com/
542 http://olderkiss.com/
543 http://chubbygirlpics.com/
544 http://gaytube.com/
545 http://juicygranny.com/
546 http://momhandjob.com/
547 http://sexybuttpics.com/
548 http://secretarypics.com/
549 http://milfkiss.com/
550 http://free-porn-pics.net/
551 http://maturedolls.net/
552 http://maturexxxclipz.com/
553 http://hairymilfpics.com/
554 http://stripping-moms.com/
555 http://pornsticky.com/
556 http://30plusgirls.com/
557 http://wifesbank.com/
558 http://sexymaturepussies.com/
559 http://zmilfs.com/
560 http://dailyolders.com/
561 http://grannyhairy.net/
562 http://7feel.net/
563 http://sexyhotmilfs.com/
564 http://milfsbeach.com/
565 http://amateur-libertins.net/
566 http://hotmomsporn.com/
567 http://milfsarea.com/
568 http://xxxmaturepost.com/
569 http://maturewitch.com/
570 http://gentlemoms.com/
571 http://posing-matures.com/
572 http://amapics.net/
573 http://matureplace.com/
574 http://wifenaked.net/
575 http://oldmomstgp.com/
576 http://agedcunts.net/
577 http://maturedummy.com/
578 http://amazingmaturesluts.com/
579 http://bigtitsfree.net/
580 http://owerotica.com/
581 http://fuckdc.com/
582 http://eroticplace.net/
583 http://fuckmaturewhore.com/

176 http://xnxx.vc/
177 http://masalopeblack.com/
178 http://porno-marocaine.com/
179 http://film-porno-black.com/
180 http://axnxxx.org/
181 http://cochonnevideosx.com/
182 http://babosas.co/
183 http://video-sex.femmesx.net/
184 http://chaudassedusexe.com/
185 http://cerdas.com/
186 http://sexe-evbony.com/
187 http://peliculaspornogratisxxx.com/
188 http://videosanalesx.com/
189 http://pornocolumbia.co/
190 http://salope-marocaine.com/
191 http://boutique-sexy.ch/
192 http://nexxx.com/
193 http://porn.com/
194 http://sexe2asiatique.com/
195 http://jeunette18.com/
196 http://puritanas.com/
197 http://les-groses.net/
198 http://beauxcul.com/
199 http://es.bravotube.net/
200 http://toroporno.com/
201 http://keezmovies.com/
202 http://pornotantique.com/
203 http://tendance-lesbienne.com/
204 http://film-xxx-black.com/
205 http://adultwork.com/
206 http://freesex.com/
207 http://porno-wife.com/
208 http://sambaporno.com/
209 http://guide-asie.com/
210 http://rubias19.com/
211 http://hairy.com/
212 http://gonzoxxxmovies.com/
213 http://dildosatisfaction.tumblr.com/
214 http://penguinvids.com/
215 http://nudevista.com/
216 http://dorceltv.xn.pl/
217 http://18teensexposed.tumblr.com/
218 http://cuckinohio.tumblr.com/
219 http://girthyencounters.tumblr.com/
220 http://gratishentai.net/
584 http://matureguide.com/
585 http://askyourmommy.com/
586 http://milffreepictures.com/
587 http://gracefulmom.com/
588 http://maturepornhere.com/
589 http://bigboty4free.com/
590 http://teenhana.com/
591 http://immaturewomen.com/
592 http://amaclips.com/
593 http://maturepicsarchive.com/
594 http://sexymaturepics.com/
595 http://tubefellas.com/
596 http://uniquesexymoms.com/
597 http://maturepornqueens.net/
598 http://tiny-cams.com/
599 http://30yomilf.com/
600 http://maturesort.com/
601 http://sex.pornoxxl.org/
602 http://everydaycams.com/
603 http://riomilf.com/
604 http://imomsex.com/
605 http://matureclits.net/
606 http://momsecstasy.com/
607 http://fresholders.com/
608 http://bizzzporno.com/
609 http://oldwomanface.com/
610 http://home-madness.com/
611 http://immodestmoms.com/
612 http://wetmaturepics.com/
613 http://gobeurettes.com/
614 http://teemns-pic.com/
615 http://worldxxxphotos.com/
616 http://old-vulva.com/
617 http://video-porno.1lecheuse.com/
618 http://maturebabesporno.com/
619 http://queenofmature.com/
620 http://hotamateurclip.com/
621 http://bigtitsporn.me/
622 http://momsinporn.net/
623 http://lewdmistress.com/
624 http://maturemomsex.com/
625 http://oldpoon.com/
626 http://posingwomen.com/
627 http://hqoldies.com/
628 http://grannypornpics.net/

221 http://stretchedpussy.tumblr.com/
222 http://femmesporno.com/
223 http://yasminramos.com/
224 http://xxxbunker.com/
225 http://whoresmilfsdegraded.tumblr.com/
226 http://bigdickswillingchicks.tumblr.com/
227 http://beeg.com/
228 http://uplust.com/
229 http://sextubelinks.com/
230 http://chaturbate.com/
231 http://theofficiallouisejenson.com/
232 http://pinkworld.com/
233 http://omegaporno.com/
234 http://tukif.com/
235 http://69rueporno.com/
236 http://indienne-sexy.com/
237 http://blogfalconstudios.com/
238 http://hard.pornoxxl.org/
239 http://arabe-sexy.com/
240 http://vivthomas.com/
241 http://ovideox.com/
242 http://xxl.sexgratuits.com/
243 http://videospornonacional.com/
244 http://youngpornvideos.com/
245 http://videos-porno.x18xxx.com/
246 http://webpnudes.com/
247 http://xxx.xxx/
248 http://smutty.com/
249 http://pornhubfillesalope.com/
250 http://teensnowxvideos.com/
251 http://store.falconstudios.com/
252 http://girlygifporn.com/
253 http://culx.org/
254 http://labatidora.net/
255 http://xbabe.com/
256 http://xxxkinky.com/
257 http://indiansexstories.net/
258 http://lookatvintage.com/
259 http://x-art.com/
260 http://bigboobsalert.com/
261 http://beautyandthebeard1.tumblr.com/
262 http://vintagehairy.net/
263 http://arabicdancevideo.blogspot.com/
264 http://mc-nudes.com/
265 http://asiatique-femme.com/
629 http://esseporn.com/
630 http://deviantclip.com/
631 http://matureinlove.net/
632 http://insext.net/
633 http://hotmomspics.com/
634 http://xxx.adulttube.com/
635 http://oldsweet.com/
636 http://amateur-sexys.tumblr.com/
637 http://nudedares.tumblr.com/
638 http://pahubad.com/
639 http://pornmirror.com/
640 http://freemilfpornpics.com/
641 http://hothomemadepix.tumblr.com/
642 http://dagay.com/
643 http://boytikol.com/
644 http://matureandgranny.com/
645 http://khu18.biz/
646 http://newsfilter.org/
647 http://sweetmaturepics.com/
648 http://adultreviews.com/
649 http://video.freex.mobl/
650 http://efukt.com/
651 http://video-one.com/
652 http://upskirt.com/
653 http://youjizz.ws/
654 http://kingsizebreasts.com/
655 http://myfreepornvideos.net/
656 http://filthyoldies.com/
657 http://myhdshop.com/
658 http://porn720.com/
659 http://ashleyrnadison.com/
660 http://bravioteens.com/
661 http://myonlyhd.com/
662 http://xxxvideo.com/
663 http://pornnakedgirls.com/
664 http://hollywoodjizz.com/
665 http://yourlust.com/
666 http://es.xhamster.com/
667 http://randyhags.com/
668 http://teenpussy.pw/
669 http://porny.com/
670 http://10pointz.com/
671 http://voyeursport.com/
672 http://dixvi.com/
673 http://pornxxx.com/

266 http://herbalviagraworld.com/
267 http://specialgays.com/
268 http://porn00.org/
269 http://gggay.com/
270 http://sexbotbonnasse.com/
271 http://megamovie.us/
272 http://salope.1japonsex.com/
273 http://redtuve.com/
274 http://kellydivine.co/
275 http://milfs30.com/
276 http://7dog.com/
277 http://onlygirlvideos.com/
278 http://ass4all.com/
279 http://freshmatureporn.com/
280 http://cindymovies.com/
281 http://roflpot.com/
282 http://maturelle.com/
283 http://dreammovies.com/
284 http://matureshine.com/
285 http://nudeboobshotpics.com/
286 http://owsmut.com/
287 http://matures-photos.com/
288 http://wetmaturewhores.com/
289 http://maturestation.com/
290 http://pornosfilms.com/
291 http://pinsex.com/
292 http://live.sugarbbw.com/
293 http://womenmaturepics.com/
294 http://hot-naked-milfs.com/
295 http://nautilix.com/
296 http://maturepornhub.com/
297 http://gay43.com/
298 http://stiflersmoms.com/
299 http://jeffdunhamfuckdoll.com/
300 http://nude-oldies.com/
301 http://liberteenage.com/
302 http://jizzle.com/
303 http://brazzersnetwork.com/
304 http://grannyxxx.co.uk/
305 http://uniquesexymoms.com/
306 http://popurls.com/
307 http://nakedboobs.net/
308 http://imaturewomen.com/
309 http://matureoracle.com/
310 http://ledauphine.com/
674 http://realitypassplus.com/
675 http://godao.com/
676 http://buzzwok.com/
677 http://sex3.com/
678 http://pornoforo.com/
679 http://freemilfsite.com/
680 http://xpornking.com/
681 http://kickass.co/
682 http://pornokutusu.com/
683 http://nakedoldbabes.com/
684 http://wixvi.com/
685 http://milfpornet.com/
686 http://adultfriendfinder.com/
687 http://pornzz.com/
688 http://kickass.com/
689 http://porntubevidz.com/
690 http://collegehumor.com/
691 http://xtube.nom.co/
692 http://playboy.com/
693 http://thehotpics.com/
694 http://daultpornvideox.com/
695 http://porndig.com/
696 http://ww.lastsexe.com/
697 http://bigtinz.com/
698 http://imagefap.com/
699 http://sexy-links.net/
700 http://hdrolet.com/
701 http://eatyouout.tumblr.com/
702 http://fakku.net/
703 http://gonzo.com/
704 http://fuck-milf.com/
705 http://es.bongacams.com/
706 http://pornstarhangout.com/
707 http://foto-erotica.es/
708 http://pornerbros.com/
709 http://barstoolsports.com/
710 http://indiangilma.com/
711 http://thehotpics.com/
712 http://masturbationaddicton.net/
713 http://motherless.com/
714 http://fr-nostradamus.com/
715 http://hardsextube.com/
716 http://sexyono.com/
717 http://japanesexxxtube.com/
718 http://thegranny.net/

311 http://milfous.com/
312 http://bitefaim.com/
313 http://nudeold.com/
314 http://mom50.com/
315 http://oldhotmoms.com/
316 http://webcam.com/
317 http://maturesinstockings.com/
318 http://riomature.com/
319 http://sexymaturethumbs.com/
320 http://hungrymatures.com/
321 http://golden-moms.com/
322 http://pandamovies.com/
323 http://teencamvids.org/
324 http://6mature9.com/
325 http://eroticbeauties.net/
326 http://multimature.com/
327 http://105matures.com/
328 http://broslingerie.com/
329 http://motherstits.com/
330 http://kissmaturesgo.com/
331 http://mulligansmilfs.com/
332 http://elderly-women.com/
333 http://upskirttop.net/
334 http://maturosexy.com/
335 http://unshavenpussies.net/
336 http://megavideoporno.org/
337 http://amateurmaturewives.com/
338 http://riomoms.com/
339 http://bestmaturewomen.com/
340 http://oldernastybitches.com/
341 http://maturewant.com/
342 http://sex.com/
343 http://riomature.com/
344 http://inlovewithboobs.com/
345 http://pornorama.com/
346 http://milfionaire.com/
347 http://momstaboo.com/
348 http://matureland.net/
349 http://momshere.com/
350 http://eros.com/
351 http://madmamas.com/
352 http://spankwire.com/
353 http://pornmaturewomen.com/
354 http://juliepost.com/
355 http://premium.gays.com/
719 http://allofteens.com/
720 http://teenpornjoy.com/
721 http://maturesensations.com/
722 http://goodgrannypics.com/
723 http://cleoteener.com/
724 http://sexyteensphotos.com/
725 http://largepontube.com/
726 http://youngmint.com/
727 http://teen18ass.com/
728 http://8nsex.com/
729 http://onlyporngif.com/
730 http://dustyporn.com/
731 http://digitalplayground.com/
732 http://youngxxxpics.com/
733 http://llveleak.com/
734 http://tinysolo.com/
735 http://bubblebuttpics.com/
736 http://mrskin.com/
737 http://anatarvasnavideos.com/
738 http://pornstarnirvna.com/
739 http://superdiosas.com/
740 http://find-best-lingerie.com/
741 http://mynakedteens.com/
742 http://pinkteenpics.com/
743 http://in.spankbang.com/
744 http://desindian.sextgem.com/
745 http://rude.com/
746 http://beeg.co/
747 http://kilopics.com/
748 http://tour.fuckmyindiangf.com/
749 http://3movs.com/
750 http://breeolson.com/
751 http://boyddl.com/
752 http://disco-girls.com/
753 http://lewd-girls.com/
754 http://ah-me.com/
755 http://porn20.org/
756 http://pinkcupid.com/
757 http://hyat.mobi/
758 http://hotsexyteensphotos.com/
759 http://9gag.tv/
760 http://freekiloclips.com/
761 http://clit7.com/
762 http://pornmdk.com/
763 http://allindiansexstories.com/

356 http://tubepornstars.com/
357 http://shemales.com/
358 http://hotnakedoldies.com/
359 http://matureandyoung.com/
360 http://muyzorras.com/
361 http://universeold.com/
362 http://mature-orgasm.com/
363 http://wetmaturewomen.com/
364 http://matureladiespics.com/
365 http://unshavengirls.net/
366 http://aztecaporno.com/
367 http://riotits.net/
368 http://womanolder.com/
369 http://bushypussies.net/
370 http://pornmaturepics.com/
371 http://hornybook.com/
372 http://ass-butt.com/
373 http://mature-galleries.org/
374 http://xixx.com/
375 http://nudematurewomenphotos.com/
376 http://toonztube.com/
377 http://primecurves.com/
378 http://arabesexy.com/
379 http://eromatures.net/
380 http://nakedbustytits.com/
381 http://watchersweb.com/
382 http://olderwomenarchive.com/
383 http://xxxmaturepost.com/
384 http://needmilf.com/
385 http://horny-matures.net/
386 http://grandmabesttube.com/
387 http://lovely-mature.net/
388 http://wild-matures.com/
389 http://mature30plus.com/
390 http://action36.com/
391 http://myfreemoms.com/
392 http://stiflersmilfs.com/
393 http://pornovideo.italy.com/
394 http://matureamour.com/
395 http://fantasticwomans.com/
396 http://lenawethole.com/
397 http://boobymilf.com/
398 http://girlmature.com/
399 http://bettermilfs.com/
400 http://themomsfucking.net/
764 http://m.chudaimaza.com/
765 http://xesi.mobi/
766 http://tubegogo.com/
767 http://iscindia.org/
768 http://fsiblog.com/
769 http://find-best-videos.com/
770 http://sexynakedamateurgirls.com/
771 http://pornodoido.com/
772 http://iloveindiansex.com/
773 http://comicmasala.com/
774 http://tubexclips.com/
775 http://jrunk.tumblr.com/
776 http://kink.com/
777 http://yehfun.com/
778 http://mega-teen.com/
779 http://haporntube.com/
780 http://tubestack.com/
781 http://yourlustgirlfriends.com/
782 http://wegret.com/
783 http://adultphonechatlines.co.uk/
784 http://jizzporntube.com/
785 http://shitbrix.com/
786 http://shitbrix.com/
787 http://hindi-sex.net/
788 http://nonvegjokes.com/
789 http://myhotsite.net/
790 http://hindiold.com/
791 http://jlobster.com/
792 http://bollywood-sex.net/
793 http://desikahani.net/
794 http://desitales.com/
795 http://pof.com/
796 http://katestube.com/
797 http://xxxsummer.net/
798 http://desikamasutra.com/
799 http://nuvid.com/
800 http://indiankahani.com/
801 http://private.com/
802 http://eternaldesire.com/
803 http://allindiansex.com/
804 http://fucking8.com/
805 http://heganporn.com/
806 http://indiansgoanal.org/
807 http://slutload.com/
808 http://desipapa.com/

| | |
|---|---|
| 401 | http://mature-women-tube.net/ |
| 402 | http://lustfuloldies.com/ |
| 403 | http://babesclub.net/ |
| 404 | http://milfatwork.net/ |
| 405 | http://oldsexybabes.net/ |
| 406 | http://nudematurespics.com/ |
| 407 | http://cocomilfs.com/ |
| 408 | http://sexymilfpussy.com/ |
| 809 | http://oigh.info/ |
| 810 | http://befuck.com/ |
| 811 | http://milfsaffair.com/ |
| 812 | http://mommyxxxmovies.com/ |
| 813 | http://matureladies.com/ |
| 814 | http://xxxolders.com/ |
| 815 | http://freeones.ch/ |

## Annexure B

## List of 144 Websites Selected through Simple Random Sampling for 200 Error Codes

| S. No. | URL | Response Time |
|---|---|---|
| 1 | http://3animalsextube.com/ | 0.000361248 |
| 2 | http://yehfun.com/ | 0.001763839 |
| 3 | http://lenawethole.com/ | 0.002221424 |
| 4 | http://grandmammapics.com/ | 0.002932147 |
| 5 | http://riomoms.com/ | 0.005340938 |
| 6 | http://smutty.com/ | 0.005459517 |
| 7 | http://live.sugarbbw.com/ | 0.006958428 |
| 8 | http://cleoteener.com/ | 0.007435106 |
| 9 | http://insext.net/ | 0.014049677 |
| 10 | http://pornmaturewomen.com/ | 0.014113661 |
| 11 | http://maturesexy.us/ | 0.016228422 |
| 12 | http://khu18.biz/ | 0.017818684 |
| 13 | http://mature-beach.com/ | 0.018080206 |
| 14 | http://playboy.com/ | 0.018912175 |
| 15 | http://sexyhotmilfs.com/ | 0.022365927 |
| 16 | http://h33t.to/ | 0.025154013 |
| 17 | http://boytikol.com/ | 0.035937881 |
| 18 | http://thehotpics.com/ | 0.039570511 |
| 19 | http://hot-naked-milfs.com/ | 0.040470587 |
| 20 | http://bestmilftube.com/ | 0.041318914 |
| 21 | http://hot-dates.info/ | 0.044223681 |
| 22 | http://7feel.net/ | 0.04702496 |
| 23 | http://riotits.net/ | 0.047327632 |
| 24 | http://bigbuttmature.com/ | 0.05116893 |
| 25 | http://shemale.asia/ | 0.052075045 |
| 26 | http://juicygranny.com/ | 0.052450123 |
| 27 | http://gorgeousladies.com/ | 0.053046929 |
| 28 | http://kirtu.com/ | 0.053348963 |
| 29 | http://hollywoodjizz.com/ | 0.059716168 |
| 30 | http://matures-photos.com/ | 0.059937952 |
| 31 | http://mypornbookmarks.com/ | 0.064390735 |
| 32 | http://allindiansexstories.com/ | 0.064553571 |
| 33 | http://thefreecamsecret.com/ | 0.06615052 |
| 34 | http://gramateurs.com/ | 0.074542667 |
| 35 | http://ladymom.com/ | 0.07710593 |
| 36 | http://wegret.com/ | 0.077963501 |
| 37 | http://amapics.net/ | 0.079099318 |
| 38 | http://xxxvideo.com/ | 0.079699115 |
| 39 | http://truthordarepics.com/ | 0.082345411 |
| 40 | http://2gayboys.com/ | 0.085998683 |
| 41 | http://watchmygf.com/ | 0.088436855 |
| 42 | http://hotmomspics.com/ | 0.089337054 |
| 43 | http://nudedares.tumblr.com/ | 0.08942667 |
| 44 | http://mamitatube.com/ | 0.090172718 |
| 45 | http://oldhotmoms.com/ | 0.095133044 |
| 46 | http://uniquesexymoms.com/ | 0.095438621 |
| 47 | http://theporndude.com/ | 0.100494219 |
| 48 | http://pornnakedgirls.com/ | 0.100877341 |
| 49 | http://multimature.com/ | 0.101555659 |
| 50 | http://sex3.com/ | 0.103007178 |
| 51 | http://inlovewithboobs.com/ | 0.103912892 |
| 52 | http://cuckinohio.tumblr.com/ | 0.104263909 |
| 53 | http://beautyandthebeard1.tumblr.com/ | 0.104890755 |
| 54 | http://matureshine.com/ | 0.105133877 |
| 55 | http://pamelapost.com/ | 0.109739157 |
| 56 | http://dorceltv.xn.pl/ | 0.109993519 |
| 57 | http://womenmaturepics.com/ | 0.111664146 |
| 58 | http://mulligansmilfs.com/ | 0.118048369 |
| 59 | http://savitabhabhi.mobi/ | 0.122294744 |
| 60 | http://momsinporn.net/ | 0.124202646 |
| 61 | http://tgpmaturewoman.com/ | 0.125595392 |
| 62 | http://oldsexybabes.net/ | 0.12628926 |
| 63 | http://momshere.com/ | 0.127566464 |
| 64 | http://matureladiespics.com/ | 0.128186013 |
| 65 | http://nudeboobshotpics.com/ | 0.130281187 |
| 66 | http://webcam.com/ | 0.131484813 |
| 67 | http://iloveindiansex.com/ | 0.1374973 |
| 68 | http://queenofmature.com/ | 0.138458178 |
| 69 | http://wetmaturewomen.com/ | 0.143510115 |
| 70 | http://milfmomspics.com/ | 0.16055026 |
| 71 | http://amaclips.com/ | 0.161505147 |
| 72 | http://teen18ass.com/ | 0.162568925 |
| 73 | http://comicmasala.com/ | 0.164401459 |
| 74 | http://freesex.com/ | 0.165651191 |
| 75 | http://drtuber.com/ | 0.167194085 |
| 76 | http://horny-olders.com/ | 0.168010173 |
| 77 | http://bigfreemature.com/ | 0.16986616 |
| 78 | http://momstaboo.com/ | 0.169903091 |
| 79 | http://motherless.com/ | 0.169943859 |
| 80 | http://matureguide.com/ | 0.170425194 |
| 81 | http://myexmilf.com/ | 0.17065674 |
| 82 | http://mature-women-tube.org/ | 0.171546047 |
| 83 | http://hdrolet.com/ | 0.171962022 |
| 84 | http://bestmatureclips.com/ | 0.172314018 |
| 85 | http://maturebabesporno.com/ | 0.173391199 |
| 86 | http://fsiblog.com/ | 0.176732368 |
| 87 | http://wetmaturewhores.com/ | 0.178879879 |
| 88 | http://rude.com/ | 0.191917485 |
| 89 | http://sexocean.com/ | 0.193969026 |
| 90 | http://thefreecamsecret.com/ | 0.195433347 |
| 91 | http://video-one.com/ | 0.195996594 |
| 92 | http://pornmdk.com/ | 0.199547059 |
| 93 | http://newsfilter.org/ | 0.200192182 |
| 94 | http://tinysolo.com/ | 0.202758822 |
| 95 | http://chubbygalls.com/ | 0.204541529 |
| 96 | http://indiansex4u.com/ | 0.207102574 |
| 97 | http://beeg.com/ | 0.207223739 |
| 98 | http://thehotpics.com/ | 0.207799146 |
| 99 | http://nude-oldies.com/ | 0.208158396 |
| 100 | http://home-madness.com/ | 0.212327204 |
| 101 | http://fuck-milf.com/ | 0.212804421 |
| 102 | http://30plusgirls.com/ | 0.214028667 |
| 103 | http://youngmint.com/ | 0.220159313 |
| 104 | http://es.bongacams.com/ | 0.223664587 |
| 105 | http://stiflersmoms.com/ | 0.224830755 |
| 106 | http://matureamour.com/ | 0.226791451 |
| 107 | http://eroticperfection.com/ | 0.228330166 |
| 108 | http://idealmilf.com/ | 0.229325861 |
| 109 | http://booloo.com/ | 0.234196577 |
| 110 | http://eatyouout.tumblr.com/ | 0.245319536 |
| 111 | http://teengayporntube.com/ | 0.245584632 |
| 112 | http://classic-moms.com/ | 0.245822184 |
| 113 | http://cocomilfs.com/ | 0.247493691 |
| 114 | http://milf-fucking.net/ | 0.250523585 |
| 115 | http://kickass.co/ | 0.250686141 |
| 116 | http://riomature.com/ | 0.252795766 |
| 117 | http://secretarypics.com/ | 0.253091235 |
| 118 | http://matureoracle.com/ | 0.258597389 |
| 119 | http://pornofemmeblack.com/ | 0.261280015 |
| 120 | http://woodstockreborn.tumblr.com/ | 0.261552105 |
| 121 | http://boobymilf.com/ | 0.261610421 |
| 122 | http://axnxxx.org/ | 0.263135305 |
| 123 | http://godao.com/ | 0.263930135 |
| 124 | http://milfgals.net/ | 0.264257914 |
| 125 | http://pinkteenpics.com/ | 0.264578326 |
| 126 | http://riomoms.com/ | 0.270511101 |
| 127 | http://sextubelinks.com/ | 0.271777567 |
| 128 | http://stiflersmilfs.com/ | 0.272107755 |
| 129 | http://jizzporntube.com/ | 0.273527917 |
| 130 | http://x-ho.com/ | 0.274659807 |
| 131 | http://sexymaturepussies.com/ | 0.277846404 |
| 132 | http://mom50.com/ | 0.302478522 |
| 133 | http://momhandjob.com/ | 0.305568364 |
| 134 | http://oldpoon.com/ | 0.307672475 |
| 135 | http://xbabe.com/ | 0.311748117 |
| 136 | http://matureclits.net/ | 0.314771352 |
| 137 | http://mega-teen.com/ | 0.314879231 |
| 138 | http://posing-matures.com/ | 0.318598041 |
| 139 | http://xhot.sextgem.com/ | 0.319080842 |
| 140 | http://sexyteensphotos.com/ | 0.320024245 |
| 141 | http://matureclithunter.com/ | 0.325844303 |
| 142 | http://xxxbunker.com/ | 0.328913749 |
| 143 | http://free-porn-pics.net/ | 0.333314408 |
| 144 | http://elderly-women.com/ | 0.333970627 |

## Annexure C

## List of 60 Websites Selected through Simple Random Sampling for 301 Error Codes

| S. No. | URL | Response Time |
|---|---|---|
| 1 | http://azgals.com/ | 0.00413422 |
| 2 | http://hqoldies.com/ | 0.004545489 |
| 3 | http://dagay.com/ | 0.009202058 |
| 4 | http://premium.gays.com/ | 0.013064506 |
| 5 | http://older-beauty.com/ | 0.019481869 |
| 6 | http://porndig.com/ | 0.023670141 |
| 7 | http://collegehumor.com/ | 0.027220661 |
| 8 | http://slutload.com/ | 0.033068573 |
| 9 | http://cumlouder.com/ | 0.035614655 |
| 10 | http://gggay.com/ | 0.041679244 |
| 11 | http://xxx.xxx/ | 0.048772664 |
| 12 | http://tukif.com/ | 0.052346999 |
| 13 | http://ixxx.ws/ | 0.068659291 |
| 14 | http://filthymamas.com/ | 0.068733354 |
| 15 | http://theofficiallouisejenson.com/ | 0.093210515 |
| 16 | http://empflix.com/ | 0.115035265 |
| 17 | http://penguinvids.com/ | 0.116084211 |
| 18 | http://indiansexstories.net/ | 0.118878238 |
| 19 | http://cliphunter.com/ | 0.119762823 |
| 20 | http://pornorc.net/ | 0.125098936 |
| 21 | http://xxxkinky.com/ | 0.133497061 |
| 22 | http://cholotube.com/ | 0.135591475 |
| 23 | http://sexxxdoll.com/ | 0.135909014 |
| 24 | http://bigboobsalert.com/ | 0.136621385 |
| 25 | http://xxx.adulttube.com/ | 0.138010787 |
| 26 | http://pinkcupid.com/ | 0.143889282 |
| 27 | http://indienne-sexy.com/ | 0.149344156 |
| 28 | http://pandamovies.com/ | 0.156900817 |
| 29 | http://imagefap.com/ | 0.157929696 |
| 30 | http://asiatique-femme.com/ | 0.158278281 |
| 31 | http://sexy-links.net/ | 0.170668197 |
| 32 | http://liberteenage.com/ | 0.172606767 |
| 33 | http://spankwire.com/ | 0.178448931 |
| 34 | http://rk.com/ | 0.196495213 |
| 35 | http://nudematurewomenphotos.com/ | 0.198808408 |
| 36 | http://myfreecams.com/ | 0.216780832 |
| 37 | http://tubestack.com/ | 0.217318215 |
| 38 | http://eros.com/ | 0.217401744 |
| 39 | http://brazzersnetwork.com/ | 0.217453978 |
| 40 | http://redtuve.com/ | 0.22708292 |
| 41 | http://find-best-videos.com/ | 0.230931956 |
| 42 | http://befuck.com/ | 0.232894675 |
| 43 | http://madmamas.com/ | 0.233737779 |
| 44 | http://shemales.com/ | 0.244824665 |
| 45 | http://pinsex.com/ | 0.247682154 |
| 46 | http://clit7.com/ | 0.247860128 |
| 47 | http://hotnakedoldies.com/ | 0.248024838 |
| 48 | http://olderkiss.com/ | 0.260667194 |
| 49 | http://indiangilma.com/ | 0.26154391 |
| 50 | http://barstoolsports.com/ | 0.266065517 |
| 51 | http://pornhub.com/ | 0.273269006 |
| 52 | http://cindymovies.com/ | 0.274257618 |
| 53 | http://mature.nl/ | 0.290020875 |
| 54 | http://breeolson.com/ | 0.305849818 |
| 55 | http://fakku.net/ | 0.308869261 |
| 56 | http://poringa.net/ | 0.310189451 |
| 57 | http://muyzorras.com/ | 0.311239653 |
| 58 | http://ledauphine.com/ | 0.324836113 |
| 59 | http://roundandbrown.com/ | 0.331503205 |
| 60 | http://videosxxxputas.xxx/ | 0.350588191 |

**Annexure D**

## List of 45 Websites Selected through Simple Random Sampling for 0/ Blank Error Codes

| S. No. | URL | Response Time |
|---|---|---|
| 1 | http://porno-marocaine.com/ | 0.010334578 |
| 2 | http://pornotantique.com/ | 0.011711055 |
| 3 | http://videosfilleschaudes.com/ | 0.01249191 |
| 4 | http://hornybook.com/ | 0.019313217 |
| 5 | http://needmilf.com/ | 0.02148685 |
| 6 | http://les-groses.net/ | 0.03371637 |
| 7 | http://cam4.in/ | 0.041456402 |
| 8 | http://gentlemoms.com/ | 0.047792515 |
| 9 | http://beurettehot.net/ | 0.04876022 |
| 10 | http://7dog.com/ | 0.053392172 |
| 11 | http://peliculaspornogratisxxx.com/ | 0.07899894 |
| 12 | http://lechecallente.com/ | 0.089320826 |
| 13 | http://tubeduporno.com/ | 0.095840535 |
| 14 | http://pornocolumbia.co/ | 0.101632043 |
| 15 | http://film-xxx-black.com/ | 0.112483044 |
| 16 | http://myhdshop.com/ | 0.115384538 |
| 17 | http://fotomujeres.pibones.com/ | 0.132646973 |
| 18 | http://bigboty4free.com/ | 0.144359354 |
| 19 | http://sexonapria.org/ | 0.151413187 |
| 20 | http://anatarvasnavideos.com/ | 0.166777846 |
| 21 | http://teemns-pic.com/ | 0.171999184 |
| 22 | http://buzzwok.com/ | 0.180479057 |
| 23 | http://grannyxxx.co.uk/ | 0.184261852 |
| 24 | http://numaturewomen.com/ | 0.186235408 |
| 25 | http://hotamateurclip.com/ | 0.196242343 |
| 26 | http://mature-galleries.org/ | 0.197018891 |
| 27 | http://sexbotbonnasse.com/ | 0.19966799 |
| 28 | http://videos-sexe.1touffe.com/ | 0.203385287 |
| 29 | http://sexcoachapp.com/ | 0.211358247 |
| 30 | http://bomnporn.com/ | 0.219040278 |
| 31 | http://sexe-evbony.com/ | 0.229066487 |
| 32 | http://fille-nue-video.com/ | 0.231289195 |
| 33 | http://myfreepornvideos.net/ | 0.234874603 |
| 34 | http://xxxsummer.net/ | 0.237960869 |
| 35 | http://hard.pornoxxl.org/ | 0.24046452 |
| 36 | http://gobeurettes.com/ | 0.245438129 |
| 37 | http://horny-matures.net/ | 0.249287809 |
| 38 | http://goodgrannypics.com/ | 0.254995314 |
| 39 | http://fapto.xxx/ | 0.264408384 |
| 40 | http://freematurevideo.net/ | 0.265120358 |
| 41 | http://videosanalesx.com/ | 0.296097021 |
| 42 | http://marocainenue.com/ | 0.301524776 |
| 43 | http://sexe2asiatique.com/ | 0.309186754 |
| 44 | http://omegaporno.com/ | 0.311849385 |
| 45 | http://sex.pornoxxl.org/ | 0.324317335 |

**Annexure E**

## List of 14 Websites Selected through Simple Random Sampling for Error Code 302

| S No. | URL | Response Time |
|---|---|---|
| 1 | http://xixx.com/ | 0.06353607 |
| 2 | http://pornstarhangout.com/ | 0.078999055 |
| 3 | http://xxxmomclips.com/ | 0.084626982 |
| 4 | http://shitbrix.com/ | 0.085224093 |
| 5 | http://agedcunts.net/ | 0.089093816 |
| 6 | http://largepontube.com/ | 0.154610763 |
| 7 | http://desikamasutra.com/ | 0.161248759 |
| 8 | http://xvideosnacional.com/ | 0.178103874 |
| 9 | http://mc-nudes.com/ | 0.194839893 |
| 10 | http://bigtitsnaked.com/ | 0.249584928 |
| 11 | http://yourlustgirlfriends.com/ | 0.258852867 |
| 12 | http://maturehere.com/ | 0.273263913 |
| 13 | http://porn.com/ | 0.303672052 |
| 14 | http://voyeursport.com/ | 0.344095134 |